\documentclass[manuscript]{aastex}

\slugcomment{\textbf{Astrophysical Journal Letters, in press}}

\shorttitle{2I/Borisov with HST}
\shortauthors{Jewitt et al.}

\begin{document}


\title{The Nucleus of Interstellar Comet 2I/Borisov}

\author{David Jewitt$^{1,2}$,
Man-To Hui$^{3}$,
Yoonyoung Kim$^4$,
Max Mutchler$^5$,
Harold Weaver$^6$ and 
Jessica Agarwal$^4$
}

\affil{$^1$ Department of Earth, Planetary and Space Sciences,
UCLA, Los Angeles, CA 90095-1567\\
$^2$ Department of Physics and Astronomy,
UCLA, Los Angeles, CA 90095-1547\\
$^3$Institute for Astronomy, University of Hawaii,  Honolulu, Hawaii 96822\\
$^4$ Max Planck Institute for Solar System Research, 37077 G\"ottingen, Germany\\
$^5$ Space Telescope Science Institute, Baltimore, MD 21218 \\
$^6$ The Johns Hopkins University Applied Physics Laboratory,  Laurel, Maryland 20723
}

\email{jewitt@ucla.edu}

\begin{abstract}
We present high resolution imaging observations of interstellar comet 2I/Borisov (formerly C/2019 Q4) obtained using the Hubble Space Telescope.  Scattering from the comet is dominated by a coma of large particles (characteristic size $\sim$0.1 mm) ejected anisotropically.  Convolution modeling of the coma surface brightness profile sets a robust  limit to the spherical-equivalent nucleus radius $r_n \le$  0.5 km (geometric albedo 0.04 assumed). We obtain an independent constraint based on the non-gravitational acceleration of the nucleus, finding $r_n >$ 0.2 km (nucleus density $\rho$ = 500 kg m$^{-3}$ assumed).  The profile and the non-gravitational constraints cannot be simultaneously satisfied if $\rho \le$ 25 kg m$^{-3}$; the nucleus of comet Borisov cannot be a low density fractal assemblage of the type proposed elsewhere for the nucleus of 1I/'Oumuamua.  We show that the spin-up timescale to outgassing torques, even at the measured low production rates, is comparable to or shorter than the residence time in the Sun's water sublimation zone. The spin angular momentum of the nucleus should be changed significantly  during the current solar fly-by.  Lastly, we find that the differential interstellar size distribution in the 0.5 mm to 100 m size range can be represented by  power laws with indices $<$ 4 and that  interstellar bodies of 100 m size scale strike Earth every one to two hundred million years.  \end{abstract}


\section{INTRODUCTION}
Comet 2I/Borisov (hereafter simply ``2I'') is the  second  interstellar interloper detected in the solar system, after 1I/'Oumuamua (``1I'').  Scientific interest in these bodies lies in their role as the first known members of an entirely new population formed, presumably, by the ejection of planetesimals in the clearing phase of external protoplanetary disks (e.g.~Moro-Martin 2018).  Curiously, the first two interstellar objects appear physically quite different. Whereas 'Oumuamua was an apparently inactive, roughly 100 m scale body with a large amplitude lightcurve (Bannister et al.~2017, Jewitt et al.~2017, Meech et al.~2017, Drahus et al.~2018)  2I more closely resembles a typical solar system comet, with  a prominent dust coma that was evident at discovery (Bolin et al.~2019, Guzik et al.~2019, Jewitt and Luu 2019) and weak spectral lines (Fitzsimmons et al.~2019, McKay et al.~2019) indicative of on-going activity.  Despite appearing inactive, comet 1I exhibited non-gravitational acceleration (Micheli et al.~2018), likely caused by recoil forces from anisotropic mass loss (although other explanations have been advanced; Bialy and Loeb 2018, Moro-Martin 2019, Flekkoy et al.~2019).  

Comet 2I passed perihelion on UT 2019 December 8 at distance $q$ = 2.006 AU and will reach Jupiter's distance in  2020 July and Saturn's by 2021 March.  Here, we present pre-perihelion Hubble Space Telescope (HST) observations at heliocentric distance $r_H$ = 2.369 AU.   A particular objective is to determine the effective size of the nucleus using the highest resolution data.

\section{OBSERVATIONS}
All observations were obtained under the Director's Discretionary Time allocation GO 16009. We used the WFC3 charge-coupled device (CCD) camera on HST, which has pixels  0.04\arcsec~wide and gives a Nyquist-sampled resolution of about 0.08\arcsec~(corresponding to 160 km at the distance of 2I).  In order to fit more exposures into the allocated time, only  half of one of two WFC3 CCDs was read-out, providing an 80" x 80" field of view, with 2I approximately centrally located.  The telescope was tracked at the instantaneous non-sidereal rate of the comet (up to about 100\arcsec~hour$^{-1}$) and also dithered to mitigate the effects of bad pixels.  All data were taken through the F350LP filter, which has peak transmission $\sim$ 28\%, an effective wavelength $\lambda_e \sim$ 5846\AA, and a full width at half maximum, FHWM $\sim$ 4758\AA.  In each of four orbits, we obtained six images each of 260 s duration.  Geometric parameters of the observations are summarized in Table (\ref{geometry}).

\section{DISCUSSION}
\noindent \textbf{Morphology:} 

\begin{figure}
\plotone{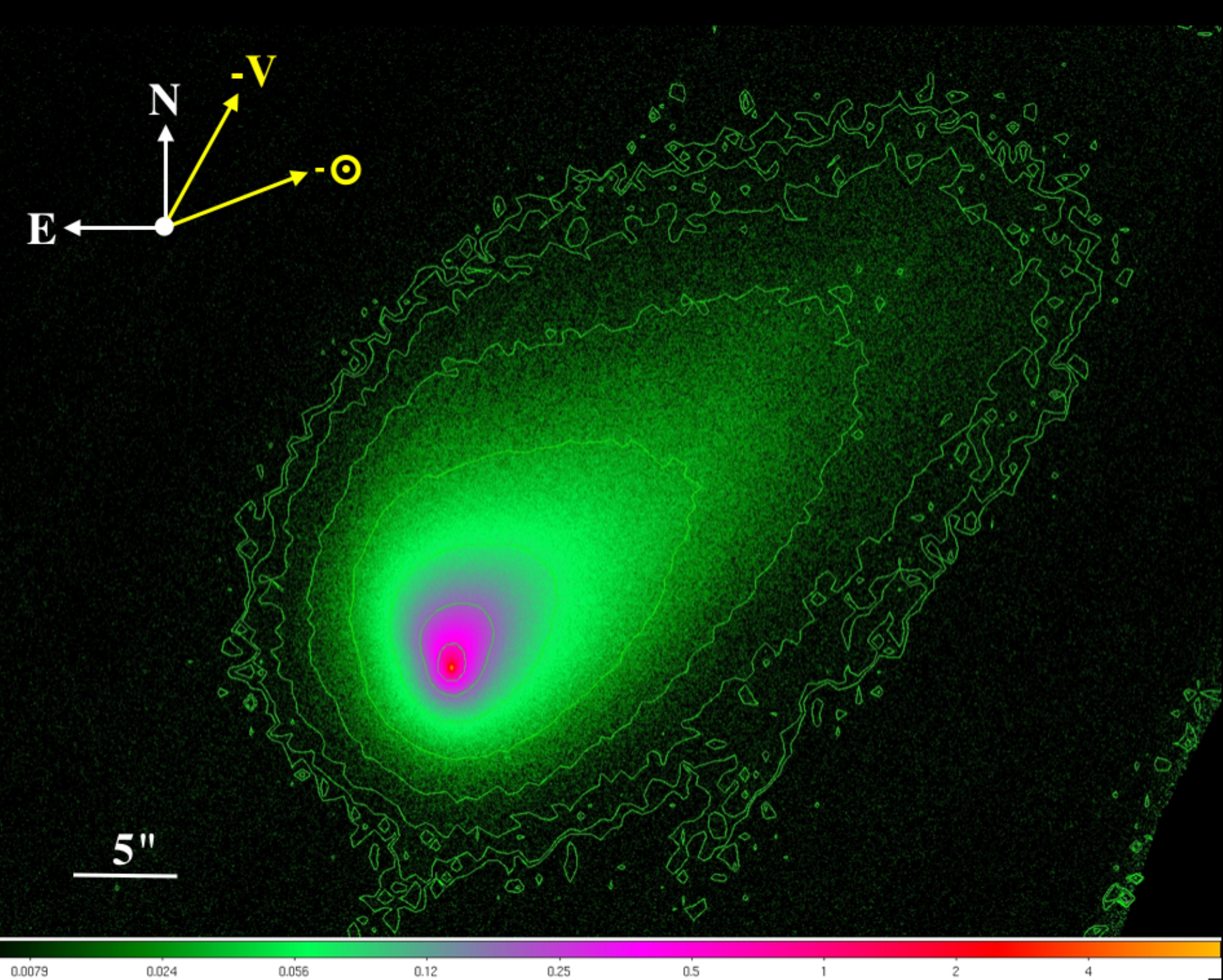}
\caption{Composite WFC3 image of comet Borisov on UT 2019 October 12 with isophotal contours overlayed. The cardinal directions are marked, as are the projected anti-solar vector ($-\odot$) and the projected negative heliocentric velocity vector ($-V$).  A 5\arcsec~(10,000 km) scale bar is shown.  \label{october12}}
\end{figure}

Figure (\ref{october12}) shows the appearance of 2I on UT 2019 October 12, in a composite image of 6240 s duration formed by aligning and combining all 24 images from the four HST orbits.  Close to the nucleus, the coma shows a bi-lobed appearance reminiscent of the morphology observed in C/2014 B1 (Jewitt et al.~2019). There, it was attributed to low latitude  ejection of large (slow) dust particles from  a nucleus rotating with its pole lying near the plane of the sky. We defer detailed modeling of the morphology of the dust coma pending the acquisition of additional images to be taken from a range of viewing perspectives.  Such observations are already planned under GO 16009 and its continuation in GO 16041.  

\begin{figure}
\plotone{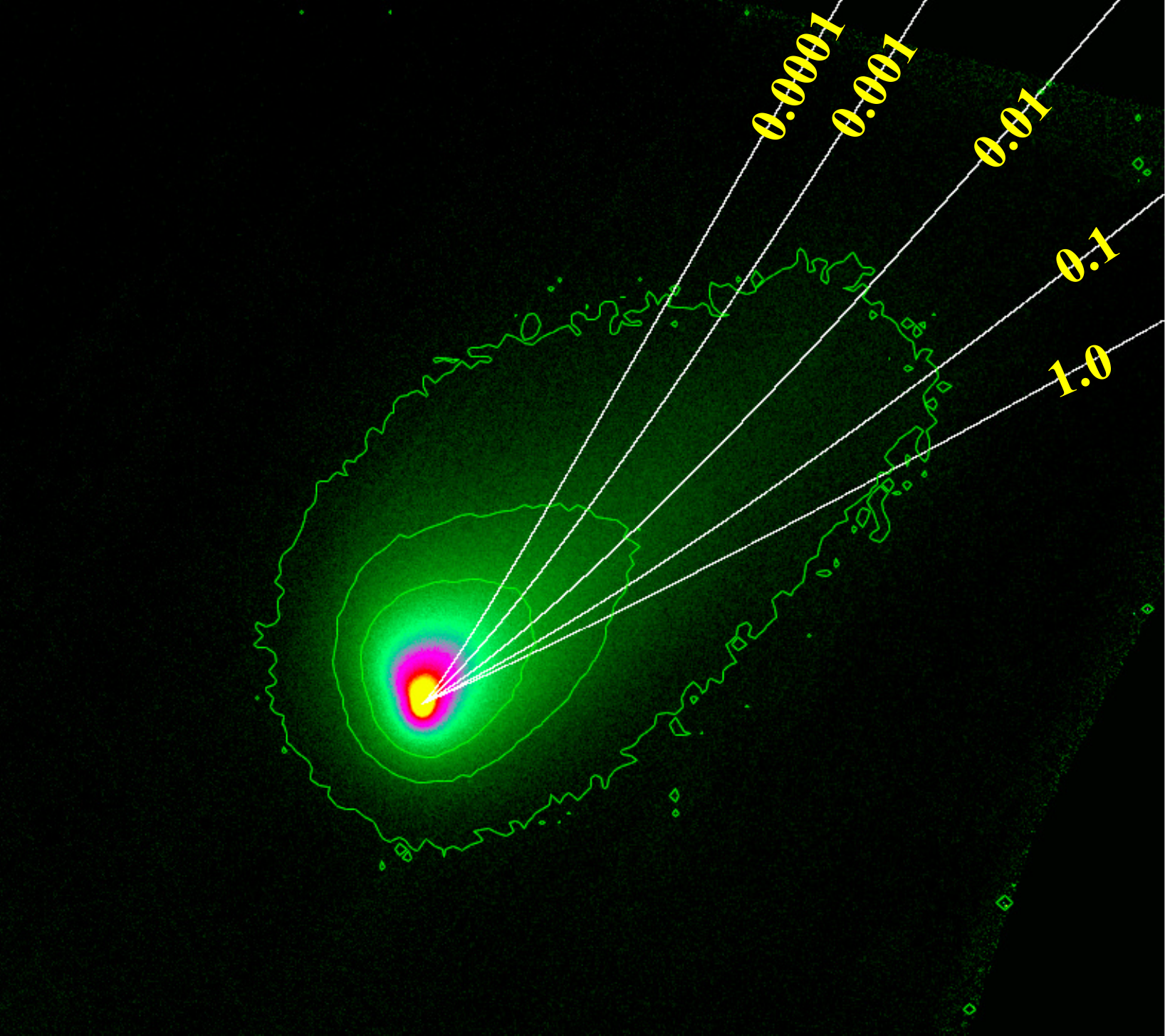}
\caption{Comet Borisov on UT 2019 October 12 showing syndynes for particles with $\beta$ = 10$^{-4}$, 10$^{-3}$, 10$^{-2}$, 10$^{-1}$ and 1, as marked.   The axis of the tail is best represented by $\beta$ = 0.01, corresponding to particle radius $a \sim$ 100 $\mu$m.  Direction arrows are the same as in Figure (\ref{october12}). \label{syndynes}}
\end{figure}

Comet 2I also shows an extensive tail of dust with an axis lying between the projected anti-solar and negative heliocentric velocity vectors, marked  in the figure by $-\odot$ and $-V$, respectively.  The tail extends to the edge of the WFC3 field of view, a distance at least 40\arcsec~from the nucleus in the composite image, corresponding to a sky-plane distance $\ell \gtrsim 8\times10^4$ km.  The tail direction in Figure (\ref{october12}) allows us to make a simple estimate of the particle size by comparison with syndyne models, shown in Figure (\ref{syndynes}).  The motion of cometary dust particles relative to their parent nucleus is controlled by the ratio of the force due to solar radiation pressure  to the gravitational force, called $\beta$.  Syndynes mark the locus of positions in the sky-plane occupied by particles having a given $\beta$ and emitted over a range of times.   The figure shows syndynes for particles with $\beta$ = 10$^{-4}$, 10$^{-3}$, 10$^{-2}$, 10$^{-1}$ and 1 ejected isotropically with zero initial relative velocity.   For dielectric particles, $\beta \sim a^{-1}$, where $a$ is the particle radius expressed in microns (Bohren and Huffman 1983).   Figure (\ref{syndynes}) shows that the tail direction is best-matched by $\beta \sim$  0.01, corresponding to particle radius $a \sim$ 100 $\mu$m (consistent with Jewitt and Luu 2019). It is clear from the complexity of the coma, particularly in the central 5\arcsec, that no single particle size model can fit the data, but we are confident that the particles are large and that micron-sized particles ($\beta$ = 1) alone cannot represent the comet.  As on other comets, the paucity of small particles might be due to cohesive forces which selectively affect the escape under gas drag.

\noindent \textbf{Photometry:} 

\begin{figure}
\plotone{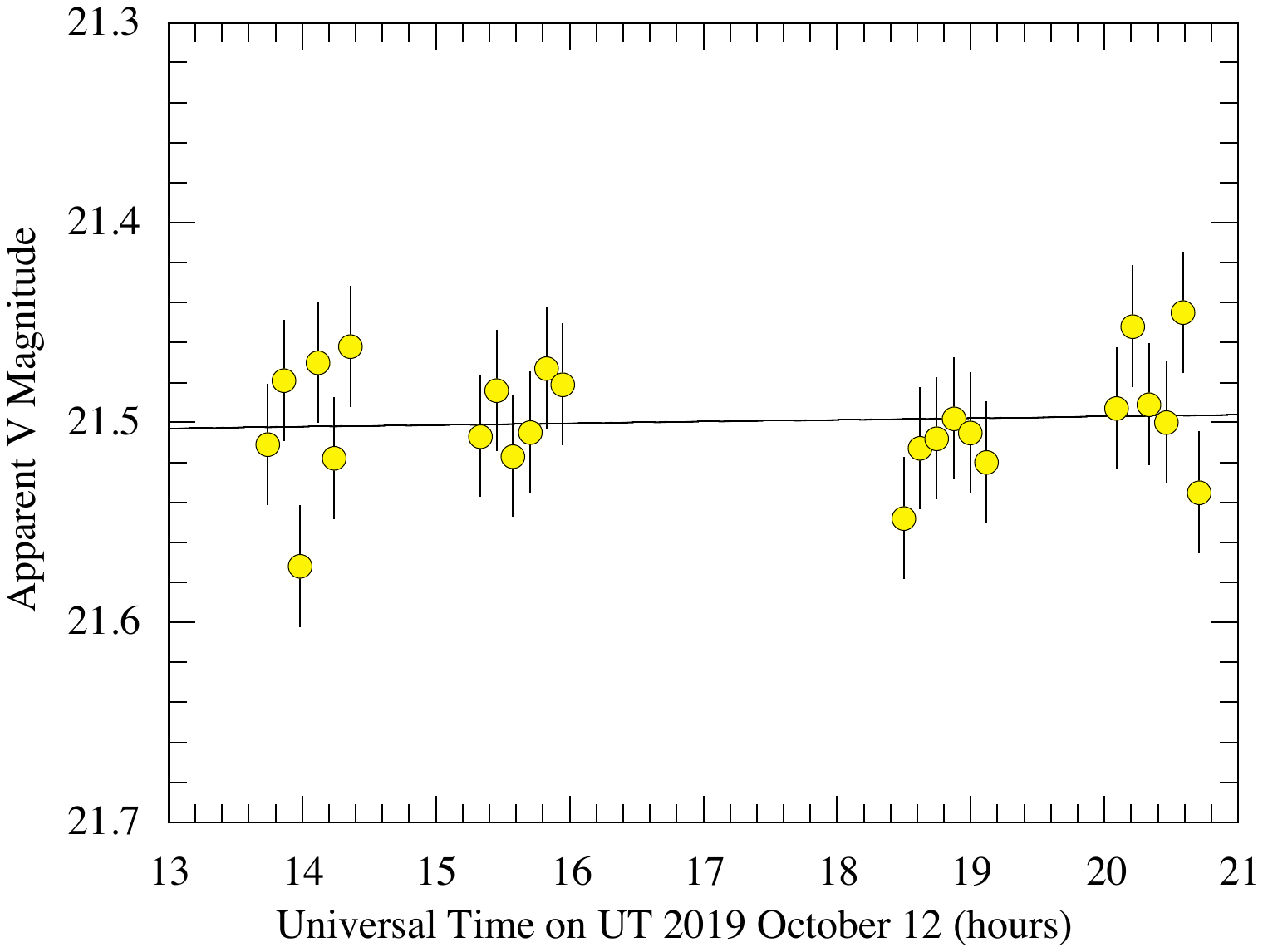}
\caption{Apparent V magnitude within a 0.2\arcsec~radius photometry aperture on UT 2019 October 12.  The straight line shows a linear, least-squares fit to the data with slope -0.001$\pm$0.002 magnitudes hour$^{-1}$. \label{lightcurve}}
\end{figure}

The time dependence of the apparent magnitude of 2I is plotted in Figure (\ref{lightcurve}), from measurements using a 0.2\arcsec~radius photometry aperture (400 km at the comet) with background subtraction from a contiguous aperture having an outer radius of 0.28\arcsec~(560 km).  The measurements span a $\sim$7 hour interval, with a gap between UT 16 and 18 hours forced by technical issues with the telescope. A linear, least-squares fit to the data (gradient $dV/dt$ = -0.001$\pm$0.002 magnitudes hour$^{-1}$), shown as a straight line in Figure (\ref{lightcurve}), indicates no systematic trend in the apparent brightness on this timescale.  However, the measured gradient is consistent with the 1\% day$^{-1}$ ($4\times10^{-4}$ magnitudes hour$^{-1}$) rate of brightening measured between September 13 and October 4  (Jewitt and Luu 2019).  Deviations from the best fit, provide no definitive evidence for short-term variations and are consistent with the 1$\sigma$ = $\pm$0.03 magnitude statistical errors.  We thus find no evidence for modulation of the  light scattered from an 'Oumuamua-like rotating, aspherical nucleus.  As we show below, the most likely explanation is coma dilution within the 0.2\arcsec~photometry aperture, because the cross-section of the nucleus is small compared to the combined cross-sections of the dust particles within the projected aperture.

\noindent \textbf{Crude Radius Estimate:}  The best-fit value of the apparent magnitude from Figure (\ref{lightcurve}) is $V$ = 21.51$\pm$0.04.   The absolute magnitude computed from $V$ assuming a linear phase function of the form $\Phi(\alpha)$ = 0.04 magnitudes degree$^{-1}$ is $H = 16.60\pm0.04$ magnitudes, where the quoted uncertainty reflects only statistical fluctuations.  Uncertainties in the phase function are systematic in nature and dominate the statistical errors.  For example, plausible uncertainties in $\Phi(\alpha)$ of $\pm$0.02 magnitudes degree$^{-1}$ affect our best estimate of $H$ by $\pm$0.4 magnitudes.

The scattering cross-section corresponding to $H$, in km$^2$, is calculated from 

\begin{equation}
C_e = \frac{1.5\times10^6}{p_V} 10^{-0.4 H}
\label{ce}
\end{equation}

\noindent where $p_V$ is the geometric albedo.  The mean albedo of cometary nuclei is $p_V \sim 0.04$ (Fernandez et al.~2013) and, using this  albedo, Equation (\ref{ce}) gives $C_e$ = 8.6 km${^2}$.  The radius of an equal-area circle is $r_e = (C_e/\pi)^{1/2}$, or $r_e$ = 1.6 km.  This constitutes a crude but absolute upper limit to the radius of the nucleus, because the central 0.2\arcsec~aperture is strongly contaminated by dust.

\begin{figure}
\epsscale{.80}
\plotone{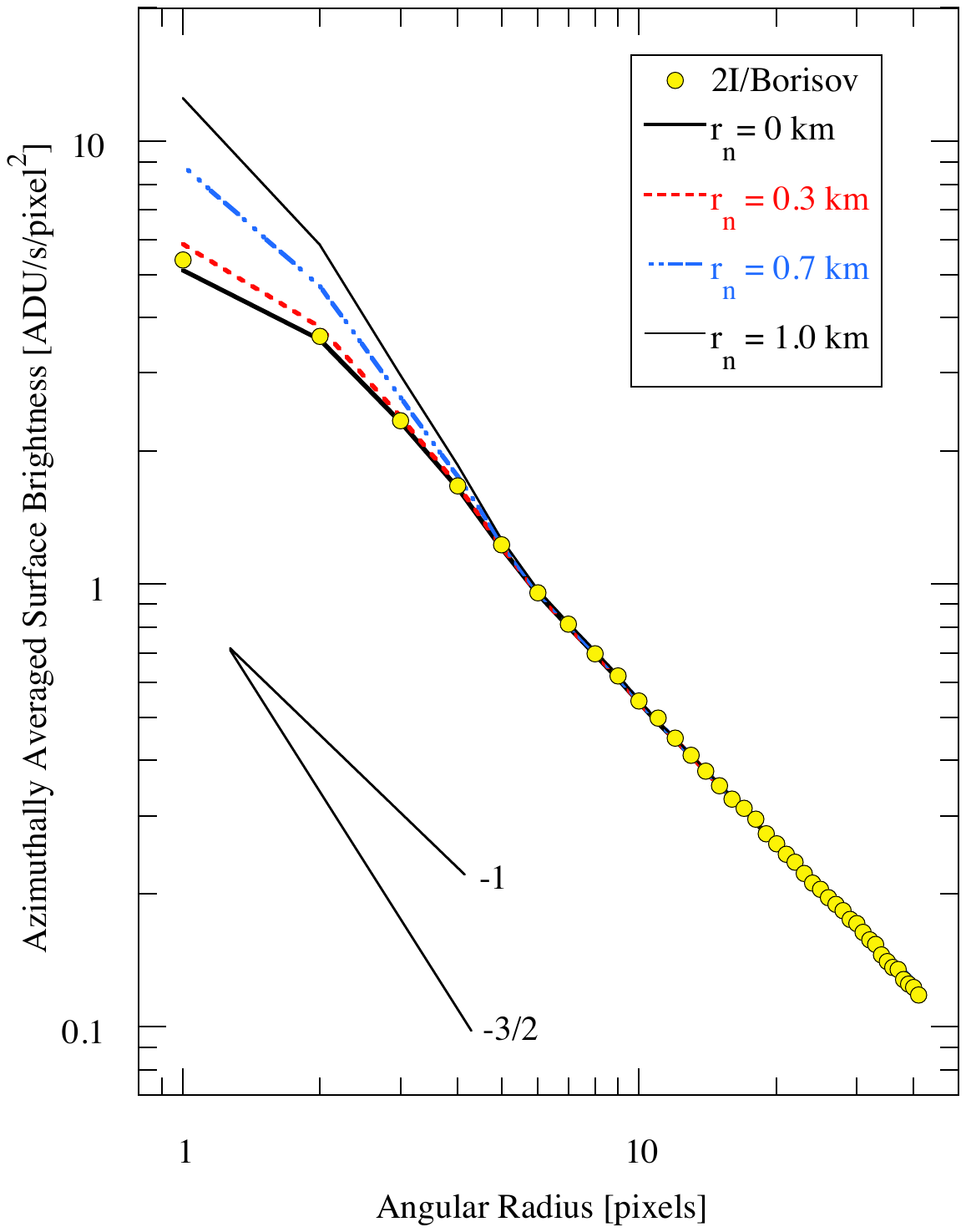}
\caption{Surface brightness profile of comet Borisov (yellow circles) compared with models in which the nucleus radius (for assumed geometric albedo $p_V$ = 0.04)  is $r_n$ = 0 km (thick black line),  0.3 km (dashed red line), 0.7 km (dash-dot blue line) and 1.0 km (thin black line).  Lines in the lower left indicate logarithmic surface brightness gradients $m$ = 1 and $m$ = -1.5, for comparison.  \label{SB_profile}}
\end{figure}

\noindent \textbf{Radius from Profile Fitting:}  To better remove the dust contamination of the near-nucleus region, we fitted a surface brightness model to the coma and extrapolated inwards to the location of the nucleus, according to the prescription described in Hui and Li (2018).  For this purpose, the surface brightness was fitted in  1\degr~azimuthal segments over the radius range 0.24\arcsec~to 1.20\arcsec.  The weighted mean surface brightness within a set of nested annuli is shown  in Figure (\ref{SB_profile}).  Statistical error bars, computed from the standard error on the mean of the value at each radius, are smaller than the plot symbols.  Systematic errors, principally due to uncertainties in the background subtraction due to field objects and internal scattering, were found to be unimportant in the fitted region of the profile. Consistent profiles were obtained using data from the other three HST orbits and so these are not shown separately.  

The fitted profiles were then  convolved with the HST point spread function provided by TinyTim (Krist et al.~2011), after adding a central point source signal to represent the nucleus.  In the figure we show models in which the  effective nucleus radii  are $r_n$ = 0, 0.3, 0.7 and 1.0 km, respectively (albedo $p_V$ = 0.04 assumed).  By inspection of the figure, we conservatively set an empirical upper limit to the radius of the nucleus  $r_n \le$ 0.5 km, far smaller than the crude estimate given above and showing the importance of accurate dust subtraction.  

\noindent \textbf{Other Radius Constraints:} Two other observations can be used to independently constrain the radius of the nucleus of 2I, although neither is as robust as the limit derived from the surface brightness profile.  Both are based on knowledge of the mass production rate of 2I.

Cometary non-gravitational motion provides a measure of the nucleus size, essentially because the sublimation rate is proportional to $r_n^2$ while the mass is proportional to $r_n^3$.  Small nuclei can be measurably accelerated by  recoil forces  from anisotropic sublimation while large nuclei cannot.  The recoil force resulting from the loss of mass at rate $\dot{M}$ and at speed $V_{th}$ is $F = f_{ng} \dot{M} V_{th}$, where $0 \le f_{ng} \le 1$ is a dimensionless number representing the degree of anisotropy of the mass loss.  Isotropic emission, corresponding to zero net force on the nucleus, has $f_{ng}$ = 0, while perfectly collimated emission has $f_{ng}$ = 1.  Observations show that sublimation-driven mass loss from comets is highly anisotropic, occurring primarily from the sun-facing hemisphere.  Through Newton's law, we set $F = 4\pi/3\ \rho r_n^3 \mathcal{A}$, where $\mathcal{A}$ is the non-gravitational acceleration [m s$^{-2}$] of the nucleus.  The effective radius of the nucleus is then given by 

\begin{equation}
r_n = \left(\frac{3 f_{ng}\dot{M} V_{th}}{4\pi \rho \mathcal{A}}\right)^{1/3}
\label{nongrav}
\end{equation}

 We use the mean thermal speed of molecules   $V_{th} = (8kT/(\pi \mu m_H))^{1/2}$, where $\mu$ = 18 is the molecular weight of H$_2$O, $m_H = 1.67\times10^{-27}$ kg is the mass of the hydrogen atom and $T$ is the temperature of the sublimating surface, which we calculated from the equilibrium sublimation condition ($T$ = 197 K at 2 AU).  We find $V_{th}$ = 480 m s$^{-1}$. The nucleus density was assumed to be $\rho$ = 500 kg m$^{-3}$ (Groussin et al.~2019). 
 
The most direct estimates of the 2I mass loss rate, $\dot{M}$, are provided by spectroscopic detections of the forbidden oxygen ([OI]6300\AA) line, which give $Q_{H2O} = (6.3\pm1.5)\times10^{26}$ ($\dot{M} = 20\pm5$ kg s$^{-1}$) at $r_H$ = 2.38 AU (McKay et al.~2019).  Less direct but broadly consistent estimates are provided by the resonant fluorescence band of CN3880\AA~(Fitzsimmons et al.~2019). When scaled to OH rates using factors determined in solar system comets (A'Hearn et al.~1995), the CN detection gives water production rates $Q_{H2O} = (1.3 - 5.1)\times10^{27}$  s$^{-1}$, corresponding to $40 \le \dot{M} \le 150$ kg s$^{-1}$ at $r_H$ = 2.66 AU.  Of these two measurements, we consider  the one based on [OI]  more likely to be accurate, given that the determination from CN is one additional step removed by the unmeasured OH/CN ratio.

 Astrometric data  provide a measure of $\mathcal{A}$, with the major limitation that, at the time of writing, the resulting deviations from purely gravitational orbit solutions are modest  (1\arcsec~to 2\arcsec), and the estimation of $\mathcal{A}$ is therefore sensitive to astrometric uncertainties as well as to the force model employed.  The solutions are particularly sensitive to pre-discovery observations reported by Ye et al.~(2019), some of which appear questionable given the low signal-to-noise ratios in their data.  
 
 The most recent available orbit solution is JPL Horizons  orbit\#48 (dated 2019 December 09), which uses astrometry obtained between UT 2018 December 13 and 2019 December 4. The solution gives $A1 = -9.1\times10^{-8}$, $A2 =2.3\times10^{-8}$ and $A3 = -1.2\times10^{-7}$.  Using the force model of Marsden et al.~(1973), the derived acceleration of the nucleus is $\mathcal{A} = 8\times10^{-7}$ m s$^{-2}$ at $r_H$ = 2 AU.  To check this, one of us (Man-To Hui) fitted a high quality sub-set of the astrometric data (enlarged to include astrometry up to 2019 December 8 ) to obtain $A1 = (7.1\pm7.1)\times10^{-8}$, $A2 = (1.0\pm1.0)\times10^{-7}$ and $A3 = (-1.6\pm1.9)\times10^{-8}$, where the quoted uncertainties are 1$\sigma$ standard deviations.  In view of the errors, the fit indicates that the three components are not statistically different from zero. Therefore, we combined the measured values in quadrature  to obtain a 3$\sigma$ limit to the acceleration at $r_H$ = 2.00 AU, $\mathcal{A} < 7\times10^{-7}$ m s$^{-2}$.  This is numerically close to the value obtained from JPL\#48, but is interpreted as an upper limit to $\mathcal{A}$ rather than a detection of it.

\begin{figure}
\epsscale{.7}
\plotone{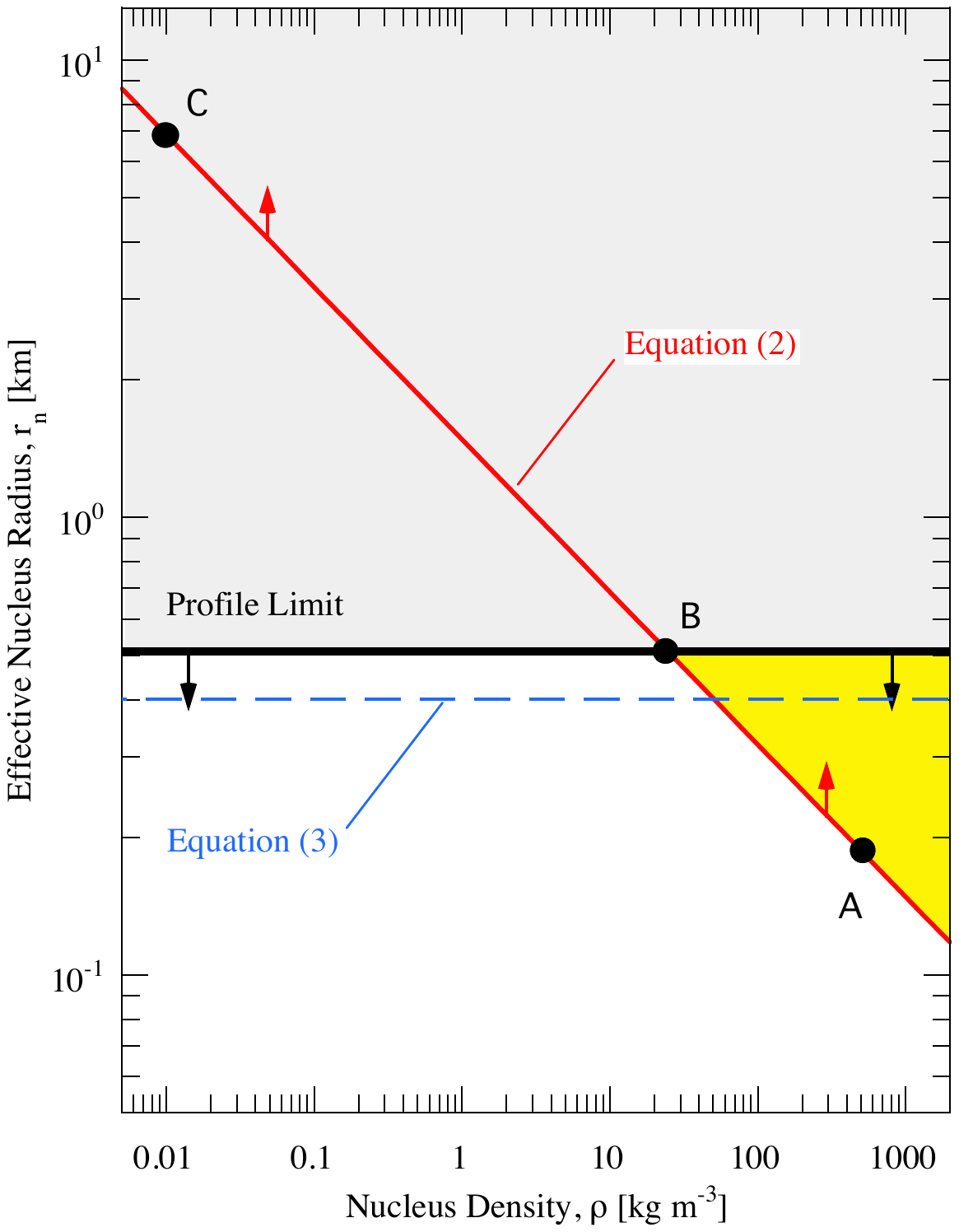}
\caption{The radius ($r_n$) vs.~nucleus density ($\rho$) plane for the nucleus of Comet Borisov.  The upper limit from the fitted profile,  $r_n < 0.5 $ km, is shown as a horizontal black line.  The non-gravitational motion limit (Equation \ref{nongrav}) is shown as a red line.    Point A shows $r_n > 0.2$ km, the radius limit inferred if the density is a comet-like $\rho$ = 500 kg m$^{-3}$.  Solutions in the ``sweet spot'' yellow triangle satisfy both the profile and non-gravitational constraints. No solutions exist for densities $\rho <$ 25 kg m$^{-3}$ (Point B).     Very low (fractal) densities like that proposed for the nucleus of 1I/'Oumuamua (Point C, with $\rho$ = 10$^{-2}$ kg m$^{-3}$) are specifically excluded. The blue dashed line shows Equation (\ref{sublimation}) computed with $f_A$ = 1.    \label{size}}
\end{figure}

With $\mathcal{A} < 7\times10^{-7}$ m s$^{-2}$, nominal density $\rho$ = 500 kg m$^{-3}$ (Groussin et al.~2019), and $f_{ng}$ = 1, Equation (\ref{nongrav}) gives $r_n >$ 0.2 km (point A in Figure \ref{size}). This value  is consistent with  $r_n <$ 0.5 km as obtained from the surface brightness profile and neatly brackets the nucleus radius, 0.2 $\le r_n \le$ 0.5 km, provided $\rho$ = 500 kg m$^{-3}$.    However, since the density of 2I is unmeasured, we must also consider other values.  In order to satisfy the profile constraint ($r_n <$ 0.5 km) while keeping $\dot{M}$ and $\mathcal{A}$ as measured, Equation (\ref{nongrav}) requires $\rho >$ 25 kg m$^{-3}$ (point B in Figure \ref{size}).  This is interesting because much lower densities  (e.g.~$\rho = 10^{-2}$ kg m$^{-3}$) have been posited in fractal models of the non-gravitational acceleration of the nucleus of 1I (Flekk{\o}y et al.~2019, Moro-Martin 2019).  Such low densities, when substituted into (Equation \ref{nongrav}) would give $r_n >$ 7 km, strongly violating the $r_n <$ 0.5 km  limit obtained from the surface brightness profile (point C in Figure \ref{size}).   

The third and weakest constraint on the nucleus radius  is based on the  production of gas. The rate of  sublimation in equilibrium with sunlight is proportional to the sublimating area according to 

\begin{equation}
r_n = \left(\frac{\dot{M}}{2\pi f_A f_s}\right)^{1/2}.
\label{sublimation}
\end{equation}

\noindent Here, $f_A > 0$ is the active fraction of the sun-facing hemisphere and $f_s$ [kg m$^{-2}$ s$^{-1}$] is the specific rate of sublimation.  We solved the equilibrium energy balance equation for water ice sublimation, neglecting conduction, assuming  heating of the sun-facing hemisphere of the nucleus.    At $r_H$ = 2.38 AU, the specific rate is $f_s = 2\times10^{-5}$ kg m$^{-2}$ s$^{-1}$, rising to $f_s = 3\times10^{-5}$ kg m$^{-2}$ s$^{-1}$ at perihelion.  Substituting active fraction $f_A$ = 1, Equation (\ref{sublimation}) gives a nominal radius $r_n$ = 0.4 km, neatly falling between the bounds set by the profile and the non-gravitational solutions (for $\rho$ = 500 kg m$^{-3}$).  However, the value of the active fraction, $f_A$, is not measured in 2I.  In well-measured solar system comets, this quantity is widely scattered from $f_A \lesssim 10^{-2}$ to $f_A \ge$ 1, with a tendency to be larger for  smaller nuclei  (A'Hearn et al.~1995).    Values $f_A < 1$ are possible if the nucleus is largely inert, allowing solutions with larger $r_n$ by Equation (\ref{sublimation}).  Values $f_A \ge$ 1 are possible when the measured gas is produced in part, or in all, by sublimation  from grains in the coma  (c.f.~Sekanina 2019), allowing solutions with smaller $r_n$.  For these reasons, while noting the amazing concordance between estimates of the nucleus radius obtained from the three different methods,  we consider the radius constraint from Equation (\ref{sublimation}) to be the weakest of the three presented here.

\subsection{Spin-Up, Size Distribution and Impact Flux}

The upper limit to the radius obtained using HST is an order of magnitude smaller than limits from ground-based data (e.g.~$r_n \lesssim$ 7 km, Ye et al.~2019).  This small radius renders the nucleus of 2I  susceptible to rapid changes in the spin state as a result of outgassing torques (Jewitt and Luu 2019).   The e-folding spin-up timescale for these torques is

\begin{equation}
\tau_s = \frac{\omega \rho r_n^4}{k_T V_{th} \dot{M}}
\label{tau}
\end{equation}

\noindent in which $\omega = 2\pi/P$ is the angular frequency of the nucleus with rotation period $P$, $\rho$ is the nucleus bulk density, $r_n$ its radius, $V_{th}$ is the velocity of material leaving the nucleus and $\dot{M}$ is the rate of loss of mass (Jewitt 1997).  The dimensionless moment arm lies in the range $0 \le k_T \le 1$, corresponding to purely  isotropic  and purely  tangential ejection, respectively.   As a guide, we cite the nucleus of 9P/Tempel 1, which had 0.005 $ \le k_T \le$ 0.04 (Belton et al.~2011). We take $P$ = 6 hours, this being the median period of nine sub-kilometer nuclei as summarized by Kokotanekova et al.~(2017).  Assuming that the bulk of the outflow momentum is carried by the gas, we again set $V_{th}$ = 480 m s$^{-1}$.  

Comet 2I spends $\sim$ 0.6 year  with $r_H <$ 3 AU, where sublimation of water is strong enough to torque the nucleus.  Setting $\tau_s$ = 0.6 year in Equation (\ref{tau}) and using the parameters listed above, we find that a nucleus smaller than $r_n \lesssim$ 0.3 ($k_T$ = 0.005) to 0.5 km ($k_T$ = 0.04) could be spun up during the current solar flyby.  This is comparable to the inferred size of 2I, 0.2 $\le r_n \le 0.5$ km, meaning that we should expect the nucleus spin to change, possibly by a large amount and conceivably enough to induce rotational breakup. Observers should be alert to this possibility when taking post-perihelion observations of the comet.

We briefly consider the implications of the discovery of 2I  (at $r_H$ = 3.0 AU) in the context of the number density of interstellar interlopers, $N_1$.   Early estimates based on the detection of 1I alone gave $N_1(r_n > 100~\mathrm{m}) \sim$ 0.1 AU$^{-3}$ (Jewitt et al.~2017) and 0.2 AU$^{-3}$ (Do et al.~2018).  At these densities we expect an average of $\sim$10 to 20  'Oumuamua scale or larger objects inside a 3 AU radius sphere at any instant (the vast majority of which are too distant and faint to be detected in on-going sky surveys). If the  number of interlopers  is distributed as a differential power law, $dN(r_n) \propto r_n^{-q} dr_n$, with index $q$ = 3 to 4 (i.e.~cumulative distribution $N(r > r_n) = \int_{r_n}^{\infty} N(r_n) dr_n \propto r_n^{1-q} \propto r_n^{-2}~\mathrm{to}~r_n^{-3}$), then the expected number of objects with $r_n \ge$ 0.5 km and $r_H < 3$ AU should fall in the range 0.8 to 0.08, respectively.  Given these expected means and assuming Poisson statistics, the probabilities of there being a single 0.5 km object  with $r_H <$ 3 AU are 0.36 and 0.07, respectively.  Therefore, the discovery of 2I is consistent with extrapolations based on 1I alone.  It might be thought that the case is complicated by  cometary activity, without which 2I would not have been discovered at small solar elongation  (in late August a bare $r_n$ = 0.5 km nucleus  would  have had apparent magnitude $V \sim$ 22.3, fainter than is reached by any near-Sun survey).   However, even without activity, the apparent magnitude of a 0.5 km nucleus in the orbit of 2I rises  to $V \sim$ 20.7 at perihelion, bright enough to be detected in several on-going sky surveys (e.g.~Catalina Sky Survey, Pan-STARRS).  We conclude that, while 2I's early detection was certainly enabled by its cometary activity, the nucleus could have been detected later even if completely inert.

Finally, we consider the impact of interstellar bodies into the Earth.  Meteor observations provide an observational constraint at small sizes.  Specifically, survey observations set an upper limit to the flux of interstellar  grains of mass $> 2\times10^{-7}$ kg (radius $>$ 0.5 mm for $\rho$ = 500 kg m$^{-3}$) of $F < 2\times10^{-4}$ km$^{-2}$ hour$^{-1}$ (Musci et al.~2012).  (Detection of a 0.5 m scale interstellar impactor has been claimed in a preprint; Siraj and Loeb 2019). The implied cumulative number density is then $N_1(>0.5 \mathrm{~mm}) = F /\Delta V$, where we take $\Delta V$ = 50 km s$^{-1}$ as the nominal impact velocity.  Substitution gives  $N_1(>0.5 \mathrm{~mm})  \lesssim 4\times10^{15}$ AU$^{-3}$.   By comparison with $N_1(>100 \mathrm{~m}) = $ 0.1 to 0.2 AU$^{-3}$, we find that the size distribution of interstellar bodies between 0.5 mm and 100 m in radius can be represented as a differential power law having  index $q < $ 4.  In such a distribution, the total mass is spread widely over a range of particle sizes, not concentrated in the smallest particles. 

At large sizes, the rate of impact of $r_n \gtrsim$ 100 m interstellar bodies into Earth (radius $R_{\oplus} = 6.4\times10^6$ m), neglecting gravitational focusing, is $\tau_I^{-1} = N_1 \pi R_{\oplus}^2 \Delta V \sim (5~\mathrm{to}~10)\times10^{-9}$ year$^{-1}$, corresponding to an impact interval 100 to 200 Myr.  Over the age of the Earth, there have been $\sim$25 to 50 such impacts, most of a scale  likely to generate airbursts.  The total mass of interstellar  material delivered is $\lesssim10^{-12} M_{\oplus}$ (1 $M_{\oplus} = 6.0\times10^{24}$ kg).  In comparison, the terrestrial collision rate with (Sun orbiting) interplanetary debris larger than 100 m is $\tau^{-1} \sim 2\times10^{-5}$ year$^{-1}$ (Equation 3 of Brown et al.~2002)  corresponding to a collision interval $\tau = 5\times10^{4}$ year.  The  ratio of the impact fluxes  from interstellar to gravitationally bound projectiles is $\tau_I/\tau \sim 10^{-4}$, indicating the relative insignificance of the former.

\clearpage 

\section{SUMMARY}
We used the Hubble Space Telescope to observe the newly-discovered  interstellar comet 2I/(2019 Q4) Borisov at the highest angular resolution.   Three independent constraints show that the nucleus is a sub-kilometer body.

\begin{enumerate}

\item  Measurements of the surface brightness profile provide the most robust (least model-dependent) constraint on the nucleus radius, $r_n \le$ 0.5 km (albedo 0.04 assumed). Substantially larger nuclei with this albedo would create a measurable excess in the central surface brightness profile that is not observed.

\item An empirical limit to the non-gravitational acceleration of the comet sets a limit to the radius  $r_n > 0.2$ km (for density $\rho$ = 500 kg m$^{-3}$).  No solution  exists for nucleus density $\rho <$ 25 kg m$^{-3}$, ruling out low density fractal structure models  as proposed elsewhere for  1I/'Oumuamua.  

\item Gas production from comet Borisov matches that expected from full surface, equilibrium sublimation of a nucleus $r_n$ = 0.4 km in radius.  However, this constraint is weak because the active fraction, $f_A$, is unmeasured;  the nucleus could be larger  if $f_A <$ 1, or smaller if  $f_A >$1, as is possible if sublimation proceeds from volatile-rich grains in the coma.

\item The spin-up time for a $r_n \le$ 0.5 km radius nucleus owing to outgassing torques  is comparable to or less than the time spent by Borisov inside 3 AU, where sublimation of water ice is non-negligible.  Therefore, we expect the spin state  to change between the discovery epoch and the final observations  in late 2020.  Rotational disruption of the nucleus might also occur.

\item The interstellar differential size distribution from 0.5 mm to 100 m  can be represented by a power law with index $q < 4$.  Interstellar objects  with $r_n \gtrsim$ 0.1 km strike Earth on average once every  10$^8$ to 2$\times10^8$ years.

\end{enumerate}

\acknowledgments

We thank Davide Farnoccia and Quanzhi Ye for discussions, Jing Li, Chien-Hsiu Lee, Amaya Moro-Martin and the anonymous referee for comments on the manuscript.  This work was supported under Space Telescope Science Institute program GO 16009.  Y.K. and J.A. were supported by the European Research Council Starting Grant 757390 ``CAstRA".

{\it Facilities:}  \facility{HST}.

\clearpage

\clearpage

\clearpage

\clearpage

\clearpage

\clearpage

\clearpage

\begin{deluxetable}{lcccrccccr}
\tablecaption{Observing Geometry 
\label{geometry}}
\tablewidth{0pt}
\tablehead{ \colhead{UT Date and Time}   & $\Delta T_p$\tablenotemark{c} & \colhead{$r_H$\tablenotemark{d}}  & \colhead{$\Delta$\tablenotemark{e}} & \colhead{$\alpha$\tablenotemark{f}}   & \colhead{$\theta_{\odot}$\tablenotemark{g}} &   \colhead{$\theta_{-V}$\tablenotemark{h}}  & \colhead{$\delta_{\oplus}$\tablenotemark{i}}   }
\startdata

2019 Oct 12  13:44 - 20:42   &  -57     &  2.369 &  2.781  & 20.4 & 292.6   & 330.1 & -13.7 \\

\enddata


\tablenotetext{a}{Airmass at the start and end time of observation}

\tablenotetext{b}{Day of Year, UT 2019 January 01 = 1}
\tablenotetext{c}{Number of days from perihelion (UT 2019-Dec-08 = DOY 342). Negative numbers indicate pre-perihelion observations.}
\tablenotetext{d}{Heliocentric distance, in AU}
\tablenotetext{e}{Geocentric distance, in AU}
\tablenotetext{f}{Phase angle, in degrees}
\tablenotetext{g}{Position angle of the projected anti-Solar direction, in degrees}
\tablenotetext{h}{Position angle of the projected negative heliocentric velocity vector, in degrees}
\tablenotetext{i}{Angle of Earth above the orbital plane, in degrees}

\end{deluxetable}

\clearpage

%
%
%
%
%
%
%
%
%
%
%
%

\clearpage


\clearpage 

%


\begin{thebibliography}{}

\bibitem[A'Hearn et al.(1995)]{1995Icar..118..223A} A'Hearn, M.~F., Millis, R.~C., Schleicher, D.~O., et al.\ 1995, \icarus, 118, 223
%
\bibitem[Bannister et al.(2017)]{2017ApJ...851L..38B} Bannister, M.~T., Schwamb, M.~E., Fraser, W.~C., et al.\ 2017, \apjl, 851, L38


\bibitem[Belton et al.(2011)]{2011Icar..213..345B} Belton, M.~J.~S., Meech, K.~J., Chesley, S., et al.\ 2011, \icarus, 213, 345



\bibitem[Bialy, \& Loeb(2018)]{2018ApJ...868L...1B} Bialy, S., \& Loeb, A.\ 2018, \apjl, 868, L1

\bibitem[Bohren, C. F., \& Huffman, D. R. 1983]{BH83} Bohren, C. F., \& Huffman, D. R. 1983, Absorption and Scattering of Light by
Small Particles (New York, Chichester, Brisbane, Toronto, Singapore: Wiley)


%
\bibitem[Bolin et al.(2019)]{2019arXiv191014004B} Bolin, B.~T., Lisse, C.~M., Kasliwal, M.~M., et al.\ 2019, arXiv e-prints, arXiv:1910.14004

\bibitem[Brown et al.(2002)]{2002Natur.420..294B} Brown, P., Spalding, R.~E., ReVelle, D.~O., et al.\ 2002, \nat, 420, 294


%
\bibitem[Borisov et al.(2019)]{2019arXiv190912144F} Borisov, G. \ 2019. Minor Planet Electronic Circular No.~2019-R106 (September 11)



%
\bibitem[Do et al.(2018)]{2018ApJ...855L..10D} Do, A., Tucker, M.~A., \& Tonry, J.\ 2018, \apjl, 855, L10
%



\bibitem[Drahus et al.(2018)]{2018NatAs...2..407D} Drahus, M., Guzik, P., Waniak, W., et al.\ 2018, Nature Astronomy, 2, 407
%
%
%
\bibitem[Fern{\'a}ndez et al.(2013)]{2013Icar..226.1138F} Fern{\'a}ndez, Y.~R., Kelley, M.~S., Lamy, P.~L., et al.\ 2013, \icarus, 226, 1138

\bibitem[Fitzsimmons et al.(2019)]{2019ApJ...885L...9F} Fitzsimmons, A., Hainaut, O., Meech, K.~J., et al.\ 2019, \apjl, 885, L9

\bibitem[Flekk{\o}y et al.(2019)]{2019ApJ...885L..41F} Flekk{\o}y, E.~G., Luu, J., \& Toussaint, R.\ 2019, \apjl, 885, L41


\bibitem[Fulle et al.(2015)]{2015ApJ...802L..12F} Fulle, M., Della Corte, V., Rotundi, A., et al.\ 2015, \apjl, 802, L12

\bibitem[Groussin et al.(2019)]{2019SSRv..215...29G} Groussin, O., Attree, N., Brouet, Y., et al.\ 2019, \ssr, 215, 29

\bibitem[Guzik et al.(2019)]{2019NatAs.tmp..467G} Guzik, P., Drahus, M., Rusek, K., et al.\ 2019, Nature Astronomy, 467


\bibitem[Hui, \& Li(2018)]{2018PASP..130j4501H} Hui, M.-T., \& Li, J.-Y.\ 2018, \pasp, 130, 104501

\bibitem[Jewitt(1997)]{1997EM&P...79...35J} Jewitt, D.\ 1997, Earth Moon and Planets, 79, 35

\bibitem[Jewitt et al.(2017)]{2017ApJ...850L..36J} Jewitt, D., Luu, J., Rajagopal, J., et al.\ 2017, \apjl, 850, L36


\bibitem[Jewitt et al.(2019)]{2019AJ....157..103J} Jewitt, D., Kim, Y., Luu, J., et al.\ 2019, \aj, 157, 103

\bibitem[Jewitt, \& Luu(2019)]{2019ApJ...886..29J} Jewitt, D., \& Luu, J.\ 2019, \apjl, 886, L29

\bibitem[Kokotanekova et al.(2017)]{2017MNRAS.471.2974K} Kokotanekova, R., Snodgrass, C., Lacerda, P., et al.\ 2017, \mnras, 471, 2974


\bibitem[Krist et al.(2011)]{2011SPIE.8127E..0JK} Krist, J.~E., Hook, R.~N., \& Stoehr, F.\ 2011, \procspie, 81270J

\bibitem[Marsden et al.(1973)]{1973AJ.....78..211M} Marsden, B.~G., Sekanina, Z., \& Yeomans, D.~K.\ 1973, \aj, 78, 211


\bibitem[McKay et al.(2019)]{2019arXiv191012785M} McKay, A.~J., Cochran, A.~L., Dello Russo, N., et al.\ 2019, arXiv e-prints, arXiv:1910.12785

\bibitem[Meech et al.(2017)]{2017Natur.552..378M} Meech, K.~J., Weryk, R., Micheli, M., et al.\ 2017, \nat, 552, 378

\bibitem[Micheli et al.(2018)]{2018Natur.559..223M} Micheli, M., Farnocchia, D., Meech, K.~J., et al.\ 2018, \nat, 559, 223

\bibitem[Moro-Mart{\'\i}n(2018)]{2018ApJ...866..131M} Moro-Mart{\'\i}n, A.\ 2018, \apj, 866, 131

\bibitem[Moro-Mart{\'\i}n(2019)]{2019ApJ...872L..32M} Moro-Mart{\'\i}n, A.\ 2019, \apjl, 872, L32

\bibitem[Musci et al.(2012)]{2012ApJ...745..161M} Musci, R., Weryk, R.~J., Brown, P., et al.\ 2012, \apj, 745, 161




\bibitem[Sekanina(2019)]{2019arXiv191106271S} Sekanina, Z.\ 2019, arXiv e-prints, arXiv:1911.06271

\bibitem[Siraj, \& Loeb(2019)]{2019arXiv190407224S} Siraj, A., \& Loeb, A.\ 2019, arXiv e-prints, arXiv:1904.07224

\bibitem[Ye et al.(2019)]{2019arXiv191105902Y} Ye, Q., Kelley, M.~S.~P., Bolin, B.~T., et al.\ 2019, arXiv e-prints, arXiv:1911.05902






\end{thebibliography}
\end{document}